\documentclass[aps,twocolumn,showpacs,10pt,superscriptaddress,preprintnumbers,nofootinbib]{revtex4-1}
\usepackage{amsmath, amssymb}
\usepackage{physics, bm}
\usepackage{graphicx}
\usepackage{xcolor, color}
\usepackage[colorlinks, linkcolor=red, citecolor=red, urlcolor=magenta]{hyperref}

\usepackage[normalem]{ulem}
\usepackage{soul}
\usepackage{multirow}
\usepackage{orcidlink}

\newcommand{\beq}[1][]{\begin{equation}\label{#1}}
\newcommand{\eeq}{\end{equation}}
\newcommand{\bea}{\begin{eqnarray}}
\newcommand{\eea}{\end{eqnarray}}
\newcommand{\nn}{\nonumber}

\begin{document}

\preprint{CPTNP-2025-035}

\title{Probing Quark Electromagnetic Properties via Entangled Quark Pairs in Fragmentation Hadrons at Lepton Colliders}

\author{Qing-Hong Cao\,\orcidlink{0000-0003-0033-2665}}
\email{qinghongcao@pku.edu.cn}
\affiliation{School of Physics, Peking University, Beijing 100871, China}
\affiliation{School of Physics, Zhengzhou University, Zhengzhou 450001, China}
\affiliation{Center for High Energy Physics, Peking University, Beijing 100871, China}

\author{Guanghui Li\,\orcidlink{0009-0001-4822-3321}}
\email{ghli@ihep.ac.cn}
\affiliation{Institute of High Energy Physics, Chinese Academy of Sciences, Beijing 100049, China}
\affiliation{School of Physical Sciences, University of Chinese Academy of Sciences, Beijing 100049, China}

\author{Xin-Kai Wen\,\orcidlink{0009-0008-2443-5320}}
\email{xinkaiwen@ihep.ac.cn (corresponding author)}
\affiliation{Institute of High Energy Physics, Chinese Academy of Sciences, Beijing 100049, China}
\affiliation{China Center of Advanced Science and Technology, Beijing 100190, China}

\author{Bin Yan\,\orcidlink{0000-0001-7515-6649}}
\email{yanbin@ihep.ac.cn (corresponding author)}
\affiliation{Institute of High Energy Physics, Chinese Academy of Sciences, Beijing 100049, China}
\affiliation{Center for High Energy Physics, Peking University, Beijing 100871, China}

\begin{abstract}
Electromagnetic dipole interactions of light quarks induce distinct spin correlations in quark pairs produced at lepton colliders, favoring entangled spin-triplet state aligned along the $\hat{z}$ axis or spin-singlet state. These correlations lead to unique $\cos(\phi_1-\phi_2)$ azimuthal asymmetries in inclusive $\pi^+\pi^-$-dihadron pair production and in back-to-back hadron pairs ($\pi\pi,K\pi,KK$), which are absent in the SM.
Using published Belle and BaBar measurements together with projected sensitivities based on ratios of azimuthal asymmetries, we demonstrate that these measurements provide significant constraints on light-quark dipole couplings, with a reduced dependence on poorly known nonperturbative fragmentation functions and free from contamination by other new physics effects. This approach offers a clean and novel probe of light-quark dipole interactions in collider experiments.
\end{abstract}

\maketitle

\emph{Introduction.---}
Color confinement in quantum chromodynamics (QCD) requires that quarks and gluons hadronize into color-neutral hadrons, a process described by fragmentation functions (FFs) that encode the probability of partons forming specific hadrons~\cite{Berman:1971xz,Field:1977fa,Feynman:1978dt}. A comprehensive understanding of these nonperturbative distributions must incorporate their dependence on both hadron and parton spin, capturing the intricate spin-orbit correlations inherent to strong interactions~\cite{Metz:2016swz}. Of particular interest is the transfer of transverse spin from a light quark to unpolarized hadrons, described by spin-dependent FFs such as collinear interference dihadron FFs (diFFs)~\cite{Collins:1993kq,Bianconi:1999cd,Barone:2001sp,Bacchetta:2011ip,Zhou:2011ba,Courtoy:2012ry,Metz:2016swz,Cocuzza:2023vqs,Cocuzza:2023oam,Huang:2024awn,Yang:2024kjn,Kang:2025zto} and transverse-momentum-dependent (TMD) Collins functions~\cite{Collins:1992kk,Kang:2015msa,Metz:2016swz,Zeng:2023nnb,Boussarie:2023izj,Zeng:2024gun,Cao:2025wfg,He:2026pxa}. These functions can be accessed through sizable $\cos(\phi_1 + \phi_2)$ azimuthal asymmetries, including the Artru-Collins asymmetry in dihadron pair production~\cite{Artru:1995zu,Boer:2003ya} and the Collins asymmetry in single-hadron pair production~\cite{Boer:1997mf,Pitonyak:2013dsu} at lepton colliders, both arising from transverse spin correlations of the quark produced in the hard scattering.

It is also well established that massless quark pairs produced in the process $e^+e^- \to \gamma^* \to q\bar{q}$ exhibit 100\% transverse spin correlation in the central scattering region, forming a maximally entangled Bell state that can enhance the observed $\cos(\phi_1 + \phi_2)$ asymmetry. Simultaneously, the spin-triplet nature of this entangled state, aligned along the $\hat{y}$ direction (normal to the scattering plane), enforces the cancellation of $\cos(\phi_1 - \phi_2)$ asymmetry~\cite{Cheng:2025cuv}. These quantum spin correlations thus serve not only as a probe of hadronization dynamics but also as a window into the spin structure of the underlying quark system.

In this Letter, we demonstrate that entangled quark pairs reconstructed from fragmentation hadrons at lepton colliders can serve as a novel probe of the electromagnetic properties of quarks, specifically their anomalous magnetic dipole moment (MDM) and electric dipole moment (EDM). 
These effects are described by flavor-diagonal dipole operators at dimension-6 in the Standard Model Effective Field Theory (SMEFT), which connect left- and right-handed quark fields and therefore provide a direct probe of chirality-violating physics beyond the Standard Model (SM). Analogous to the anomalous magnetic moment of the muon, dipole interactions are highly sensitive to short-distance chirality-flipping dynamics. For light quarks, however, confinement prevents a direct determination of the corresponding dipole moments, making high-energy observables an essential probe of these interactions. Such operators require a chirality flip and arise naturally in ultraviolet completions containing heavy states coupling to both quark chiralities, as well as in strongly-coupled composite scenarios, constituting a characteristic low-energy effective manifestation of the underlying dynamics. Consequently, light-quark dipole operators provide a distinctive probe of a broad class of chirality-violating new physics (NP) scenarios~\cite{daSilvaAlmeida:2019cbr,Cao:2021trr,Boughezal:2023ooo,Wen:2023xxc,Shao:2023bga,Li:2024iyj,Wen:2024cfu,Wen:2024nff,Shao:2025xwp,Li:2025fom,Huang:2025ljp}.
We demonstrate for the first time that these chirality-violating interactions lead to distinct spin structures: MDM interactions produce quark pairs in an entangled spin-triplet state aligned with the $\hat{z}$ axis, while EDM interactions preferentially generate a spin-singlet configuration. These novel spin correlations generate sizable $\cos(\phi_1 - \phi_2)$ azimuthal asymmetry in inclusive $\pi^+\pi^-$ dihadron pair production and in back-to-back hadron pairs ($\pi\pi,K\pi,KK$), which are absent in standard QCD processes. 
Unlike conventional searches based primarily on inclusive production rates, the observables considered here directly probe the spin structure of the produced quark-antiquark pair through experimentally accessible azimuthal correlations. This observation establishes a direct connection between quantum entanglement, QCD spin correlations in hadronization, and probes of physics beyond the SM. We therefore suggest revisiting existing data from Belle and BaBar to search for potential signatures of quark electromagnetic dipole interactions. These measurements could provide significant constraints on quark dipole interactions through a characteristic correlation pattern that is uniquely generated by dipole operators and not by dimension-6 or dimension-8 four-fermion operators, thereby providing information complementary to conventional collider probes such as Drell-Yan production and potentially serving as a useful input to future global SMEFT analyses.

\emph{Spin correlation and quark dipole moments.---}
The general spin density matrix of a $q\bar{q}$ system can be parameterized as
\begin{equation}\label{eq:rhoCij}
\rho = \frac{ I_2 \otimes I_2 + B_i\sigma_i \otimes I_2 + \bar{B}_i I_2 \otimes \sigma_i + C_{ij} \sigma_i \otimes \sigma_j}{4},
\end{equation}
where $I_2$ is the 2-dimensional identity matrix, and $\sigma_i$ are the Pauli matrices. For light quarks produced at lepton colliders, it is convenient to adopt the helicity basis in the center-of-mass frame of the $q\bar{q}$ pair, where the $\hat{z}$-axis aligns with the quark momentum and the transverse spin components lie in the ($\hat{x}$, $\hat{y}$)-plane. The polarization vectors $B_i$ ($\bar{B}_i$) encode the spin information of the individual quark (antiquark) and have been utilized to probe signals of NP~\cite{Wen:2024cfu,Wen:2024nff}. The spin correlation tensor $C_{ij}$, widely studied in the context of quantum entanglement~\cite{Cheng:2025cuv}, also provides collider observables sensitive to nontrivial spin dynamics. In particular, the linear combinations $\mathcal{B}_{\pm} \equiv C_{xx} \pm C_{yy}$ have been employed as Bell-type variables for testing violations of the Bell inequality~\cite{Cheng:2025cuv}.

Within the SM, the spin correlation matrix for the process $e^+ e^- \to \gamma^* \to q\bar{q}$ at leading order is given by
\begin{equation}\label{eq:Cij_SM}
    C_{ij} = \mathrm{diag}\left( \frac{\sin^2\theta}{1+\cos^2\theta},-\frac{\sin^2\theta}{1+\cos^2\theta},1 \right),
\end{equation}
where $\theta$ is the hard scattering angle. This shows that the quark and antiquark are transversely correlated along $\hat{x}$-direction and anti-correlated along $\hat{y}$-direction, reaching maximal correlation at $\theta=\pi/2$, corresponding to a spin-triplet Bell state along $\hat{y}$-axis. From Eq.~\eqref{eq:Cij_SM}, one finds $\mathcal{B}_+=0$ and $\mathcal{B}_{-}=2\sin^2\theta/(1+\cos^2\theta)$, thus, any measured nonzero $B_+$ would provide a clean and sensitive signal of NP, essentially free from SM contamination.

To explore possible deviations in a model-independent framework, we adopt the SMEFT, which encodes the effects of ultraviolet (UV) dynamics through higher-dimensional operators~\cite{Buchmuller:1985jz,Grzadkowski:2010es}. Although numerous SMEFT operators up to $\mathcal{O}(1/\Lambda^4)$, with $\Lambda$ denoting the NP scale, can contribute to the process $e^+ e^- \to q\bar{q}$, only the quark dipole interactions can induce a nontrivial contribution to the observable $\mathcal{B}_+$. This is because dipole-induced production preferentially yields either an entangled spin-triplet state aligned along the $\hat{z}$ axis or a spin-singlet state, both leading to quark spin correlation patterns that are distinct from the SM.

The effective Lagrangian for quark dipole interactions relevant to our analysis is given by 
\begin{align}
\mathcal{L}_{\rm eff}&=\bar q_L\sigma^{\mu\nu}(C^{u}_BB_{\mu\nu}+C^{u}_W\tau^IW_{\mu\nu}^I)\frac{\widetilde{H} }{\Lambda^2}u_R\nn\\
&+\bar q_L\sigma^{\mu\nu}(C^d_{B}B_{\mu\nu}+C^d_{W}\tau^IW_{\mu\nu}^I)\frac{H}{\Lambda^2}d_R+\mathrm{h.c.},
\end{align}
where $q_L$, $u_R$ and $d_R$ denote the left-handed quark doublet and right-handed up- and down-type quark fields, respectively. The field strength tensors $B_{\mu\nu}$ and $W^I_{\mu\nu}$ correspond to the $U(1)_Y$ and $SU(2)_L$ gauge groups, and $H$ is the Higgs doublet with vacuum expectation value $v = 246\,\mathrm{GeV}$. At energies well below the $Z$-boson mass, as in the Belle and BaBar experiments, only the photon dipole couplings remain relevant after electroweak symmetry breaking:
\begin{align}
\mathcal{L}_{\rm eff}\supset \frac{v}{\sqrt{2}\Lambda^2}\left(C^u_{\gamma}\bar{u}_L\sigma^{\mu\nu}u_R+C^d_{\gamma}\bar{d}_L\sigma^{\mu\nu}d_R\right)A_{\mu\nu}+\mathrm{h.c.},\nn
\end{align}
where the photon dipole coefficients are given by $C^{u,d}_\gamma =\pm s_W C^{u,d}_W + c_W C^{u,d}_B$, with the sign ``$\pm$" referring to the up- and down-type quarks, and $s_W = \sin\theta_W$, $c_W = \cos\theta_W$, where $\theta_W$ is the weak mixing angle.  A calculation of the spin-correlation matrix arising purely from the dipole interactions shows that the quark-antiquark pair is produced in the entangled state $\frac{1}{\sqrt{2}}(\ket{\uparrow_{z}\downarrow_{z}}+e^{i\phi}\ket{\downarrow_{z}\uparrow_{z}})$, where $\ket{\uparrow_{z}}$ and $\ket{\downarrow_{z}}$ denote spin eigenstates along the $\hat{z}$ axis. Because a $CP$ transformation interchanges the spins of quark and antiquark, the relative phase is fixed: $\phi=0$ for a real (CP-even) dipole coupling and $\phi=\pi$ for an imaginary (CP-odd) one. Consequently, the real part yields an entangled spin-triplet state aligned along $\hat{z}$, with correlation matrix $C_{ij}^{\text{Re}}=\mathrm{diag}(1,1,-1)$, while the imaginary part produces a spin-singlet state characterized by $C_{ij}^{\text{Im}}=\mathrm{diag}(-1,-1,-1)$. This distinctive pattern makes $\mathcal{B}_+$ an especially sensitive probe of dipole-induced NP.

Focusing on the transverse spin correlations, the $C_{ij}$ components with $i,j=x,y$, evaluated up to $\mathcal{O}(1/\Lambda^4)$, are
\begin{equation}\label{eq:Cij_NP}
    \begin{aligned}
        &C_{xx}+C_{yy}  = \frac{\sin^2\theta}{1+\cos^2\theta}\frac{s v^2}{\pi\alpha \Lambda^4 Q_q^2}([{\rm Re}C_\gamma^q]^2-[{\rm Im}C_\gamma^q]^2), \\
        &C_{xx}-C_{yy}  = \frac{\sin^2\theta}{1+\cos^2\theta} (2 + \frac{1}{\Lambda^2}\mathcal{F}_1 + \frac{1}{\Lambda^4}\mathcal{F}_2), \\
        &C_{xy}= -C_{yx}  = - \frac{\sin^2\theta}{1+\cos^2\theta} \frac{s v^2}{\pi\alpha \Lambda^4 Q_q^2} [{\rm Re}C_\gamma^{q}] [{\rm Im}C_\gamma^{q}], 
    \end{aligned}
\end{equation} 
where $Q_q$ is the quark electric charge, $\alpha$ is the fine-structure constant, and $\mathcal{F}_{1,2}$ are polynomials of the coefficients of dimension-6 and dimension-8 four-fermion operators. These operators contribute only to $C_{xx}-C_{yy}$, whereas $C_{xx}+C_{yy}$ and $C_{xy,yx}$ vanish in the SM and arise solely from the dipole operators. The latter correlations therefore provide a distinctive spin-correlation signature of dipole interactions, allowing them to be isolated from generic four-fermion SMEFT contributions. The relative sign between $[\mathrm{Re}C_\gamma^q]^2$ and $[\mathrm{Im}C_\gamma^q]^2$ encodes the production of entangled triplet versus singlet states, while EDM-induced $CP$ violation then enforces $C_{xy}=-C_{yx}$. Although NP may also induce net polarization vectors $B_i(\bar B_i)$, these do not affect the Bell-type observables and are thus neglected in our analysis.

\begin{figure}
    \centering
    \includegraphics[width=0.494\linewidth]{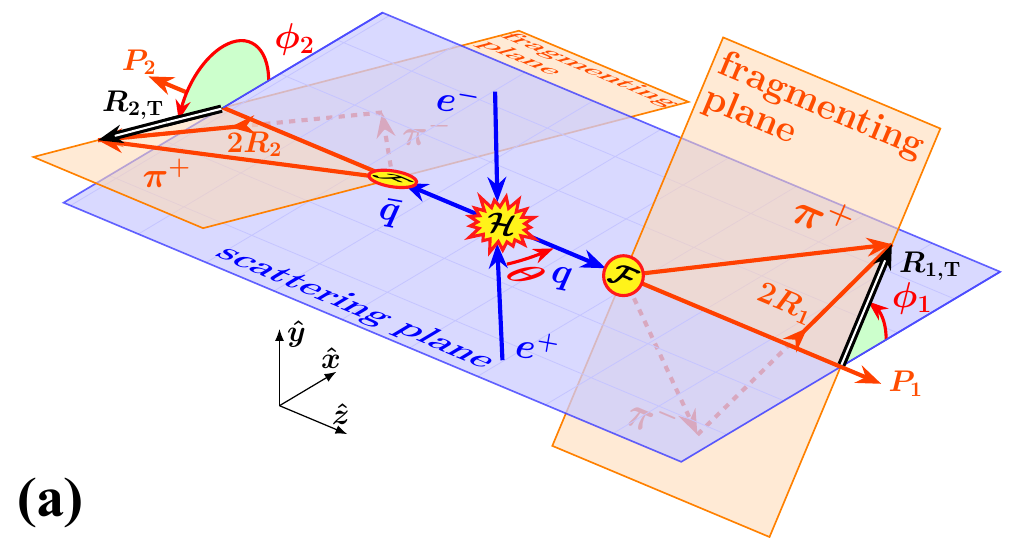}
    \includegraphics[width=0.494\linewidth]{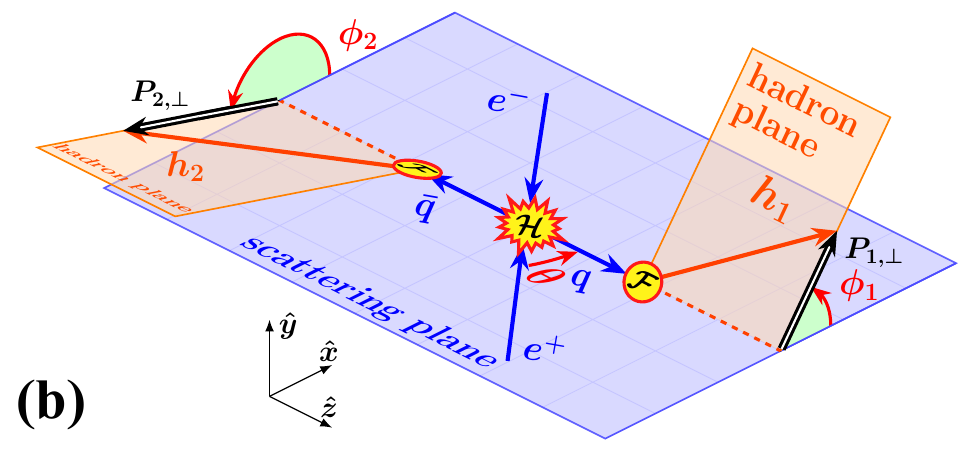}
    \caption{Leading order kinematic configuration of (a) $\pi^+\pi^-$ dihadron pair production and (b)  $h_1h_2$ hadron pair production at lepton colliders. A local coordinate system is defined with the $\hat{z}$-axis along the thrust axis, $\hat{y} = \bm{\ell} \times \hat{z} / |\bm{\ell} \times \hat{z}|$, and $\hat{x} = \hat{y} \times \hat{z}$, forming the transverse plane.}
    \label{fig:geometry}
\end{figure}

\emph{Fragmentation of quark pair and observables.---}
We begin with the semi-inclusive process $e^{-}(\ell) e^{+}(\ell^\prime) \to q(k_1)\bar{q}(k_2) \to (\pi^+\pi^-)+(\pi^+\pi^-)+X$, whose leading-order kinematics is illustrated in Fig.~\ref{fig:geometry}(a), with the $\hat{z}$-axis chosen as the thrust axis. The kinematics of each $\pi^+\pi^-$ pair are characterized by their total momentum $P_i = p_i^{\pi^+} + p_i^{\pi^-}$ and relative momentum $R_i = (p_i^{\pi^+} - p_i^{\pi^-})/2$, for $i = 1,2$. In the collinear factorization framework, assuming $|\mathbf{P}_i| \gg M_i$, the differential cross section for the $\pi^+\pi^-$ pair final states can be expressed in terms of the invariant masses $M_i^2 \equiv P_i^2$, the azimuthal angles $\phi_i$ of the transverse component $\vec{R}_{i,T}$, and the light-cone momentum fractions $z_1 = P_1^+/k_1^+$, $z_2 = P_2^-/k_2^-$, where $P_i^\pm$ and $k_i^\pm$ denote the light-cone components of $P_i$ and $k_i$, respectively~\cite{Collins:1993kq,Collins:2011zzd},
\begin{align}\label{eq:diFF}
    &\frac{1}{\sigma_0}\frac{\mathrm{d} \sigma(e^+e^-\to (\pi^+\pi^-)(\pi^+\pi^-)X)}{\mathrm{d}z_1\mathrm{d}z_2 \mathrm{d}M_1 \mathrm{d}M_2  \mathrm{d}\phi_1\mathrm{d}\phi_2\mathrm{d}\cos\theta}\nn\\
    &=(1+\cos^2\theta) \Bigg[ \sum_q Q_q^2 D_1^{q}(z_1,M_1)D_1^{\bar q}(z_2,M_2) \nn\\
    &+   \frac{1}{2} \sum_q Q_q^2  H_1^{\sphericalangle,q}(z_1, M_1) H_1^{\sphericalangle,\bar q}(z_2, M_2) 
    \Big(\mathcal{B}_- \cos(\phi_1+\phi_2)\nn\\
    &-\mathcal{B}_+ \cos(\phi_1-\phi_2)+\mathcal{B}^\prime_+ \sin(\phi_1+\phi_2)+\mathcal{B}^\prime_- \sin(\phi_1-\phi_2)\Big)\Bigg], 
\end{align}
where $\mathcal{B}_\pm = C_{xx} \pm C_{yy}$, $\mathcal{B}^\prime_\pm = C_{xy} \pm C_{yx}$ and $\sigma_0=3\pi\alpha^2/(2s)$. Here, $D_1^q(z_i,M_i)$ and $H_1^{\sphericalangle,q}(z_i, M_i)$ denote the unpolarized and interference diFFs, respectively.
From Eq.~\eqref{eq:Cij_NP}, we find $\mathcal{B}^\prime_+ = 0$, while  $\mathcal{B}^\prime_-\neq 0$ provides a clean signature of $CP$ violation from a quark EDM. 
To fully capture the dynamics, one needs to consider the contributions from both quarks and antiquarks through their respective fragmentation functions.
It is worth noting that when the thrust axis is used to define the kinematics, ambiguities in assigning the quark and antiquark directions cause the $\sin(\phi_1-\phi_2)$ distribution to average to zero over symmetric final states. In principle, this $CP$-violating effects could still be probed in processes where the quark and antiquark directions are distinguishable, such as hadron-in-jet measurements with jet charge identification~\cite{Li:2019dre,Kang:2020fka,Li:2021uww,Kang:2021ryr,Li:2023tcr,Kang:2023ptt,Wang:2023azz,Cui:2023kqb}. Therefore, for the inclusive $\pi^+\pi^-$ dihadron pair production, only the $CP$-even observables $\mathcal{B}_\pm$   are experimentally accessible. 
These can be extracted from the azimuthal modulations as
\begin{align}\label{eq:Bpm}
A_\pm &\equiv 2\langle \cos(\phi_1 \pm \phi_2) \rangle\nn\\
&=\pm\frac{\sum_q Q_q^2 H_1^{\sphericalangle,q}(z_1,M_1)H_1^{\sphericalangle,\bar q}(z_2,M_2)\mathcal{B}_\mp}{2\sum_q Q_q^2 D_1^q(z_1,M_1) D_1^{\bar q}(z_2,M_2)},
\end{align}
where $A_\pm$ denotes the experimentally observed azimuthal asymmetry, measured from the average value of $\cos(\phi_1\pm\phi_2)$. Notably, $\mathcal{B}_+$ exhibits a nontrivial dependence on quark flavor induced by NP, as shown in Eq.~\eqref{eq:Cij_NP}.
To eliminate dependence on these nonperturbative diFFs, we define the ratio
\begin{align}\label{eq:RD}
R_{\rm diFF} = \frac{A_-}{A_+}&=-\frac{\sum_q Q_q^2 H_1^{\sphericalangle,q}(z_1,M_1)H_1^{\sphericalangle,\bar q}(z_2,M_2)\mathcal{B}_+}{\sum_q Q_q^2 H_1^{\sphericalangle,q}(z_1,M_1)H_1^{\sphericalangle,\bar q}(z_2,M_2)\mathcal{B}_-}\nn\\
&\simeq-\frac{s v^2}{2\pi\alpha \Lambda^4}\frac{\sum_{q=u,d}([{\rm Re}C_\gamma^q]^2-[{\rm Im}C_\gamma^q]^2)}{\sum_{q=u,d} Q_q^2},
\end{align}
which isolates underlying spin correlation structure. Here, we have used the flavor symmetries of the interference diFFs implied by isospin and charge conjugation: $H_1^{\sphericalangle,u}=-H_1^{\sphericalangle,d}$, $H_1^{\sphericalangle,s,\bar{s}, c,\bar{c}, b, \bar{b}}=0$, and  $H_1^{\sphericalangle,q}=-H_1^{\sphericalangle,\bar{q}}$~\cite{Bacchetta:2011ip,Cocuzza:2023vqs}.
While these symmetry relations need not hold exactly, possible deviations would induce residual corrections to Eq.~\eqref{eq:RD}. Nevertheless, the dominant diFF dependence cancels in the ratio, resulting in a substantially reduced sensitivity to nonperturbative fragmentation functions compared with the individual asymmetries.

The transverse spin correlation can similarly be accessed in $e^+e^-\to q\bar{q} \to h_1h_2 +X$ within the TMD framework,  with leading order kinematics shown in Fig.~\ref{fig:geometry}(b) 
where the $\hat{z}$-axis is aligned with the thrust direction as before. The corresponding differential cross section reads~\cite{Boer:1997mf,DAlesio:2007bjf}
\begin{align}\label{eq:Collins}
    &\frac{1}{\sigma_0}\frac{\mathrm{d} \sigma(e^+e^-\to h_1h_2X)}{ p_{1,\perp}p_{2,\perp}\mathrm{d}z_1\mathrm{d}z_2 \mathrm{d}p_{1,\perp} \mathrm{d}p_{2,\perp}  \mathrm{d}\phi_1\mathrm{d}\phi_2\mathrm{d}\cos\theta}\nn\\
    &=(1+\cos^2\theta)\Bigg[  \sum_q Q_q^2 D_1^{q}(z_1,p_{1,\perp})D_1^{\bar q}(z_2,p_{2,\perp}) \nn\\
    &+   \frac{1}{2} \sum_q Q_q^2 \mathcal{F}_h H_1^{\perp,q}(z_1, p_{1,\perp}) H_1^{\perp,\bar q}(z_2,p_{2,\perp})
    \Big(\mathcal{B}_- \cos(\phi_1+\phi_2)\nn\\
    &-\mathcal{B}_+ \cos(\phi_1-\phi_2)+\mathcal{B}^\prime_+ \sin(\phi_1+\phi_2)+\mathcal{B}^\prime_- \sin(\phi_1-\phi_2)\Big)\Bigg], 
\end{align}
where $\mathcal{F}_h=(p_{1,\perp}p_{2,\perp})/(z_1z_2M_{h_1}M_{h_2})$, with $p_{i,\perp}$ and $M_{h_i}$ denoting the transverse momentum and mass of hadron $h_i$. Here, $D_1(z_i,p_{i,\perp})$ and $H_1^{\perp, q}(z_i,p_{i,\perp})$ are the unpolarized TMD FF and Collins function, respectively. The momentum fractions $z_{1,2}$ are defined analogously to the dihadron pair production case. The azimuthal modulations again allow extraction of $\mathcal{B}_\pm$, from which we define the ratio $R_{\rm Collins}=A_-/A_+$, with results similar to Eq.~\eqref{eq:Bpm} and Eq.~\eqref{eq:RD}.
For $h_1h_2=\pi\pi$ or $KK$, the TMD FFs cancel in the ratio $R_{\rm Collins}$, as in Eq.~\eqref{eq:RD}, whereas for $K\pi$ final states the cancellation is incomplete, leaving a residual FFs dependence that must be taken into account when extracting the underlying spin correlations. This arises because the Collins function distinguishes between favored and unfavored fragmentation, where ``favored'' refers to hadrons containing the fragmenting quark as a valence quark, whereas ``unfavored'' refers to hadrons that do not~\cite{Anselmino:2007fs,Zeng:2023nnb}. While $KK$ final states can also constrain strange-quark dipole couplings, their statistical precision is limited compared to light ($u,d$) quarks. For these reasons, our numerical analysis focuses primarily on $\pi\pi$ final states.

To reduce the impact of detector acceptance and radiative effects in extracting Collins asymmetries for $h_1h_2$ hadron pair production, both Belle and BaBar employ ratios of normalized distributions from different data samples. Specifically, they analyze the ratio of unlike-sign pairs ($h_1^\pm h_2^\mp$, denoted U) to like-sign pairs ($h_1^\pm h_2^\pm$, denoted L), and the ratio of unlike-sign to any-charge combinations (denoted C). As an illustration, the normalized ratio can be derived directly from Eq.~\eqref{eq:Collins},
\begin{align}
\frac{N_U}{N_L}\simeq 1+\sum_q \mathcal{F}_{UL}^q\left[\mathcal{B}_- \cos(\phi_1+\phi_2)-\mathcal{B}_+ \cos(\phi_1-\phi_2)\right],
\end{align} 
where $\mathcal{F}_{UL}^q=\alpha_{p_{1\perp},p_{2\perp}}^{z_1,z_2}(U)-\alpha_{p_{1\perp},p_{2\perp}}^{z_1,z_2}(L)$. The event-by-event factor $\alpha_{p_{1\perp},p_{2\perp}}^{z_1,z_2}$ is given by, 
\beq
\alpha_{p_{1\perp},p_{2\perp}}^{z_1,z_2}=\mathcal{F}_h\frac{ Q_q^2 H_1^{\perp,q}(z_1,p_{1,\perp})H_1^{\perp,\bar q}(z_2,p_{2,\perp})}{2\sum_q Q_q^2 D_1^q(z_1,p_{1,\perp}) D_1^{\bar q}(z_2,p_{2,\perp })}.
\eeq
Similarly, one can define the normalized ratio $N_U/N_C$. From these ratios, the corresponding observables $R_{\rm Collins}^{UL}$ and $R_{\rm Collins}^{UC}$ are obtained by evaluating the average value of $\cos(\phi_1\pm \phi_2)$ with different data samples. We present these results separately, as they are derived from the same underlying data sets.

\emph{Estimated sensitivity and discussion.---}
With the method developed above, we reinterpret existing measurements of $A_+$ from inclusive $\pi^+\pi^-$ dihadron pair production at Belle~\cite{Belle:2011cur}, and from $h_1h_2=\pi\pi$ production at Belle~\cite{Belle:2008fdv} and BaBar~\cite{BaBar:2013jdt,BaBar:2015mcn}, to constrain quark dipole interactions. Since measurements of $A_-$ are not currently available, we assume that the binning for $A_-$ matches that of $A_+$, with identical statistical and systematic uncertainties. This allows us to construct the expected ratio observables $R_{\rm diFF}$ and $R_{\rm Collins}^{UC,UL}$.

We then perform a $\chi^2$ analysis over all bins from the $\pi^+\pi^-$ dihadron pair and $h_1h_2=\pi\pi$ measurements at Belle and BaBar to constrain the quark dipole interactions,
\begin{equation}
\chi^2 = \sum_i \frac{\left(R_{\rm th}^i - R_{\rm exp}^i\right)^2}{\delta R_i^2},
\end{equation}
where $R_{\rm th}^i$ and $R_{\rm exp}^i$ denote the SMEFT theoretical predictions and experimental measurements of the ratio observables $R_{\rm diFF}$ and $R_{\rm Collins}^{UC,UL}$, respectively. We take $R_{\rm exp}^i = 0$ in accordance with the SM expectation, and 
include only statistical uncertainties $\delta R_i$, assuming that many systematic uncertainties cancel in the ratios due to the common event samples and analysis procedures.

\begin{figure}
    \centering
    \includegraphics[width=0.494\linewidth]{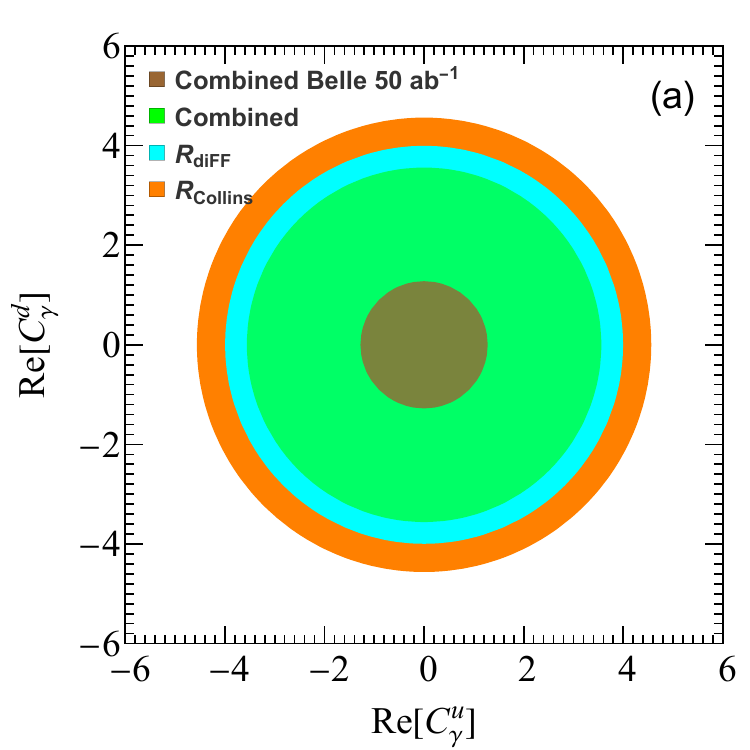}
    \includegraphics[width=0.494\linewidth]{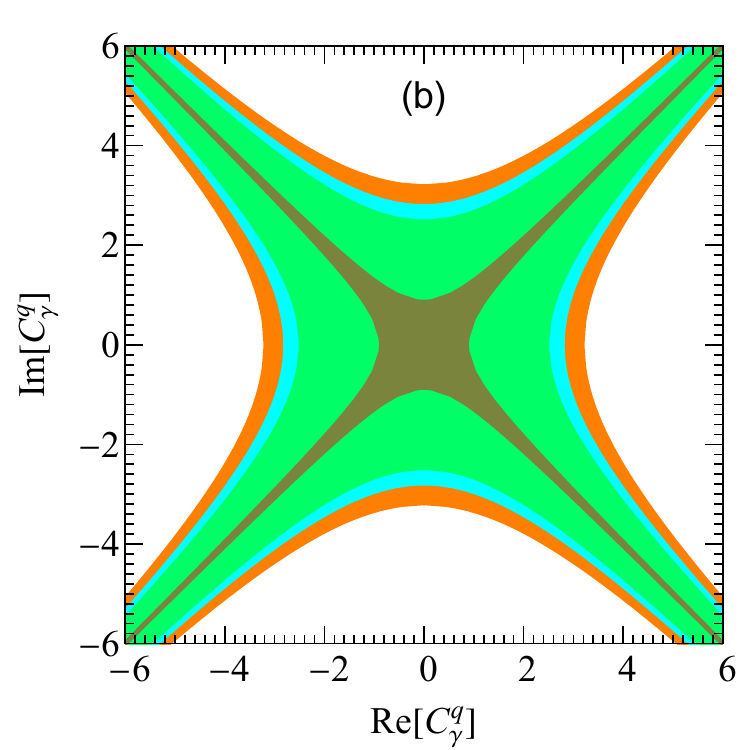}
    \caption{Expected sensitivity to the electromagnetic dipole couplings from $R_{\rm Collins}$ (orange), $R_{\rm diFF}$ (cyan), their combination (green), and the Belle upgrade with $\mathcal{L}=50~{\rm ab}^{-1}$ (brown), assuming (a) $CP$ conservation and (b) quark flavor universality.}
    \label{fig:limits}
\end{figure}

Figure~\ref{fig:limits} shows the resulting sensitivity to the electromagnetic dipole couplings of light quarks at the benchmark scale $\Lambda = 1~{\rm TeV}$ using the standard criterion $\Delta \chi^2 = 1$, from $R_{\rm diFF}$ (cyan), $R_{\rm Collins}^{UL}$ (orange), and their combination (green) using Belle and BaBar data, 
together with the projected sensitivity from the Belle upgrade at $\mathcal{L}=50~{\rm ab}^{-1}$ (brown). For illustration, results are presented under two benchmark scenarios: (a) $CP$ conservation and (b) quark flavor universality ($C_\gamma^u=C_\gamma^d\equiv C_\gamma^q$). Current data yield $\mathcal{O}(1)$ bounds on the up- and down-quark couplings in the $CP$-conserving case, identical to those for $CP$ violation since Bell-type observables are insensitive to the $CP$ nature of the couplings and the limits are dominated by $R_{\rm diFF}$. Results from $R_{\rm Collins}^{UC}$ are comparable and therefore not shown.

Electromagnetic dipole couplings can also be constrained by neutron EDM measurements and by high-energy Drell-Yan production. Under single-operator assumptions, neutron EDM bounds place extremely stringent constraints on the CP-violating components of the dipole couplings~\cite{Kley:2021yhn}, while Drell-Yan measurements away from the $Z$ pole provide strong collider constraints on the dipole operators~\cite{Alioli:2018ljm,Boughezal:2021tih,Grunwald:2023nli}. The corresponding Drell--Yan limits obtained in single-operator fits are at the level of $\mathcal{O}(0.1)$. However, neither constraint is fully model independent. Neutron EDM observables receive contributions from multiple CP-violating operators, while global SMEFT analyses of Drell-Yan processes including additional dimension-6 and dimension-8 operators could weaken the bounds on individual coefficients by more than an order of magnitude due to degeneracies among EFT directions~\cite{Boughezal:2021tih}. In contrast, the transverse spin-correlation observables considered here are generated uniquely by the light-quark dipole operators and are not induced by all the dimension-6 and dimension-8 four-fermion operators considered in this work. They therefore provide complementary information that is free from these operator degeneracies.

\emph{Conclusions.---}
In this Letter, we demonstrate that the electromagnetic dipole interactions of light quarks drive quark pairs produced at lepton colliders into either entangled spin-triplet state aligned along the $\hat{z}$ axis or spin-singlet state, yielding spin correlation patterns distinct from the SM.
Consequently, they generate unique $\cos(\phi_1-\phi_2)$ azimuthal asymmetries in inclusive $\pi^+\pi^-$ dihadron pair production and in back-to-back hadron pairs ($\pi\pi,K\pi,KK$) at lepton colliders, which are absent in the SM. We therefore propose revisiting existing Belle and BaBar measurements, together with projected sensitivities for the corresponding $\cos(\phi_1+\phi_2)$ asymmetries, within both the collinear and TMD factorization frameworks to probe light-quark dipole couplings. By considering the ratios of azimuthal asymmetries $\cos(\phi_1-\phi_2)$ and $\cos(\phi_1+\phi_2)$, we show that the sensitivity to nonperturbative fragmentation functions is reduced. This provides significant constraints on quark dipole couplings that are free from contamination by other possible NP effects in the hard scattering process, making the approach complementary to conventional probes such as neutron EDM measurements and Drell-Yan production at the LHC.

\vspace{3mm}
The authors thank Kun Cheng and Zhite Yu for helpful discussion. The work is partly supported by the National Natural Science Foundation of China under Grant Nos.~12547174, ~12422506, ~12342502, ~12235001 and CAS under Grant No.~E429A6M1, and is partly supported by Fundamental and Interdisciplinary Disciplines Breakthrough Plan of the Ministry of Education of China-JYB2025XDXM204 and by the China Postdoctoral Science Foundation under Grant No.~2026M793746. The authors gratefully acknowledge the valuable discussions and insights provided by the members of the Collaboration on Precision Tests and New Physics (CPTNP).

\bibliographystyle{apsrev}
\bibliography{reference}

\end{document}